\documentclass[12pt,preprint]{aastex}
\newcommand{\Mearth}{$M_\oplus$}

\usepackage{amsmath}
\usepackage{wasysym}
\usepackage{amsmath}
\usepackage{amssymb}
\usepackage{color}
\usepackage{amstext}
\usepackage{longtable}
\usepackage{ulem}
\usepackage{graphicx}
\usepackage{epstopdf}
\usepackage{epsfig}
\usepackage{multirow}

\usepackage{color}
\definecolor{myColor}{rgb}{0.9,0.9,0.9}  

\begin{document}
\shorttitle{New insights on Saturn's formation from its nitrogen isotopic composition}
\shortauthors{O. Mousis et al.}
\title{New insights on Saturn's formation from its nitrogen isotopic composition \\%[5cm]
}


 \author{Olivier~Mousis\altaffilmark{1,2}, Jonathan~I.~Lunine\altaffilmark{2}, Leigh~N.~Fletcher\altaffilmark{3}, Kathleen~E.~Mandt\altaffilmark{4}, Mohamad~Ali-Dib\altaffilmark{5}, Daniel~Gautier\altaffilmark{6}, and Sushil~Atreya\altaffilmark{7}}
\altaffiltext{1}{Aix Marseille Universit{\'e}, CNRS, LAM (Laboratoire d'Astrophysique de Marseille) UMR 7326, 13388, Marseille, France {\tt olivier.mousis@lam.fr}}
\altaffiltext{2}{Center for Radiophysics and Space Research, Space Sciences Building Cornell University,  Ithaca, NY 14853, USA}
\altaffiltext{3}{Atmospheric, Oceanic and Planetary Physics, Clarendon Laboratory, Parks Road, Oxford, OX1 3PU, UK}
\altaffiltext{4}{Southwest Research Institute, San Antonio, TX 78228, USA}
\altaffiltext{5}{Universit{\'e} de Franche-Comt{\'e}, Institut UTINAM, CNRS/INSU, UMR 6213, Observatoire des Sciences de l'Univers de Besan\c con, France}
\altaffiltext{6}{LESIA, Observatoire de Paris, CNRS, UPMC, Univ. Paris-Diderot, France}
\altaffiltext{7}{Department of Atmospheric, Oceanic, and Space Sciences, University of Michigan, USA}

\begin{abstract}

The recent derivation of a lower limit for the $^{14}$N/$^{15}$N ratio in Saturn's ammonia, which is found to be consistent with the Jovian value, prompted us to revise models of Saturn's formation using as constraints the supersolar abundances of heavy elements measured in its atmosphere. Here we find that it is possible to account for both Saturn's chemical and isotopic compositions if one assumes the formation of its building blocks at $\sim$45 K in the protosolar nebula, provided that the O abundance was $\sim$2.6 times protosolar in its feeding zone. To do so, we used a statistical thermodynamic model to investigate the composition of the clathrate phase that formed during the cooling of the protosolar nebula and from which the building blocks of Saturn were agglomerated. We find that Saturn's O/H is at least $\sim$34.9 times protosolar and that the corresponding mass of heavy elements ($\sim$43.1 \Mearth) is within the range predicted by semi-convective interior models.

\end{abstract}



\section{Introduction}

The measurements of $^{14}$N/$^{15}$N ratios throughout the solar system can be divided into three categories (see \cite{2014ApJ...788L..24M} and references therein): the solar wind and Jupiter have the lightest ratios, presumed to be representative of the protosolar ratio. Chondrites, grains coming from comet 81P/Wild 2, Earth's mantle and atmosphere, Venus and Mars' mantle have moderately heavy ratios. Saturn's moon Titan, Mars' atmosphere, as well as NH$_3$ and HCN in comets share the lowest $^{14}$N/$^{15}$N values.

The recent derivation of a 1--sigma lower limit for the $^{14}$N/$^{15}$N ratio in Saturn's ammonia, which is found to be $\sim$500 from TEXES/IRTF ground-based mid-infrared spectroscopic observations \citep{2014Icar..238..170F}, prompts us to revise models of Saturn's formation that previously only used the supersolar abundances of heavy elements measured in the observable troposphere as constraints. This lower limit is formally consistent with the $^{14}$N/$^{15}$N ratio ($\sim$435) measured by the Galileo probe at Jupiter \citep{2014Icar..238..170F} and implies that the two giant planets were essentially formed from the same nitrogen reservoir in the nebula, which is N$_2$ \citep{2001ApJ...553L..77O,2014Icar..238..170F}. Any scenario depicting Saturn's formation should match the $^{14}$N/$^{15}$N ratio measured in its atmosphere and be consistent with disk's temperatures greater than 30 K in the giant planets formation region. Lower temperatures have only been observed in regions located beyond $\sim$30 AU in circumstellar disks \citep{2013Sci...341..630Q}.

Two scenarios of Saturn's formation, aiming at matching the supersolar volatile abundances measured in its envelope, have been proposed. Both approaches determine the composition of the planet's building blocks from a simple clathrate formation model and assume that all elements were in protosolar abundances in the disk's gas phase. The first scenario, proposed by \cite{2008P&SS...56.1103H}, assumes that Saturn formed at $\sim$40--50K in the protosolar nebula (hereafter PSN). In their model, NH$_3$ was trapped in planetesimals, while the dominant N molecule in the PSN, N$_2$, remained well mixed with H$_2$ until the gas collapsed onto the core of the planet. This scenario is now ruled out because it suggests that Saturn's supersolar N abundance essentially results from the delivery of NH$_3$ trapped in solids, implying that its $^{14}$N/$^{15}$N ratio should be substantially lower than the Jovian value \citep{2008P&SS...56.1103H}.

Alternatively, \cite{2009ApJ...696.1348M} proposed that Saturn's building blocks formed at a cooler temperature in the disk. In this scenario, planetesimals were agglomerated from a mixture of clathrates and pure ices condensed close to $\sim$20 K, implying that both NH$_3$ and N$_2$ were trapped in solids. Their model is consistent with the measured $^{14}$N/$^{15}$N ratio since N$_2$ remains the main nitrogen reservoir delivered to Saturn. However, the formation of Saturn's building blocks at such a low temperature in the PSN is questionable as the heating of the disk by proto-Sun's UV radiation might prevent the temperature from decreasing down to $\sim$40 K at 10 AU \citep{1998ApJ...500..411D}.

Here we find that it is possible to account for both Saturn's chemical and isotopic compositions if one assumes the formation of its building blocks at $\sim$45 K in the PSN, provided that the O abundance was $\sim$2.6 times protosolar in its feeding zone. To do so, we used a statistical thermodynamic model (Mousis et al. 2010, 2012) to investigate the composition of the clathrate phase that formed during the cooling of the PSN from the most abundant gaseous volatiles. These clathrates agglomerated with the other ices and rocks and formed the building blocks of Saturn. A fraction of these planetesimals accreted in the growing Saturn dissolved in its envelope and subsequently engendered the observed volatile enrichments.

\section{Useful elemental abundances measured in Saturn}

Table 1 summarizes the abundances of C, N, P, S and O, normalized to their protosolar abundances, and measured in the forms of CH$_4$, NH$_3$, PH$_3$, H$_2$S (indirect determination) and H$_2$O in Saturn's atmosphere. Note that the protosolar abundances correspond to the present day solar values corrected from elemental settling in the Sun over the past 4.56 Gyr \citep{2009LanB...4B...44L}. The abundance of CH$_4$ has been determined from the analysis of high spectral resolution observations from Cassini/CIRS \citep{2009Icar..199..351F}. As methane does not condense at Saturn's atmospheric temperatures, its atmospheric abundance can be considered as representative of the bulk interior. The NH$_3$ abundance is taken from the range of values derived at the equator by \cite{2011Icar..214..510F} from Cassini/VIMS 4.6--5.1 $\mu$m thermal emission spectroscopy. The measured NH$_3$ abundance may be considered as a lower limit since the condensation level of NH$_3$--bearing volatiles may be deeper than the sampled regions \citep{1999P&SS...47.1243A,2014P&SS...47.1243A}, implying that there could be a large reservoir of ammonia hidden below the condensate cloud decks. PH$_3$ has been determined from Cassini/CIRS observations at 10 $\mu$m \citep{2009Icar..202..543F}. The H$_2$S abundance is quoted from the indirect determination of \cite{1989Icar...80...77B} from radio observations but remains highly uncertain.

\section{Model description}

In our model, the volatile phase incorporated in planetesimals is composed of a mixture of pure ices, stoichiometric hydrates (such as NH$_3$--H$_2$O hydrate) and multiple guest (MG) clathrates\footnote{A MG clathrate forms from a mixture of several gases and is consequently occupied simultaneously by several species.} that crystallized in the form of microscopic grains at various temperatures in the outer part of the disk. We assume that planetesimals have grown from collisional coagulation of the icy grains \citep{1997Icar..127..290W}. Here, the clathration process stops once crystalline water ice has been consumed by the trapping of volatile species in clathrate and hydrate phases. Only pure condensates can subsequently form if the disk cools down to very low temperatures. The process of volatile trapping in planetesimals formed in Saturn's feeding zone follows the approach depicted in \cite{2012ApJ...757..146M} who used a statistical thermodynamic model to compute the composition of MG clathrates formed in the PSN. We refer the reader to this paper for further information on the employed model. We use typical PSN temperature and pressure profiles \citep{hg} to compute the evolution of the thermodynamic conditions at the current location of Saturn. Here, our computations have been made in the case of formation of Structure I MG clathrates because CO, CO$_2$ and H$_2$S, namely the most abundant volatiles in the PSN, also individually form Structure I clathrates.

Our computations are based on a predefined initial gaseous composition in which all elemental abundances, except that of oxygen in some circumstances (see Sec. \ref{discu}), are protosolar \citep{2009LanB...4B...44L}. We assume that O, C, and N exist only under the form of H$_2$O, CO, CO$_2$, CH$_3$OH, CH$_4$, N$_2$, and NH$_3$. Hence, once the gaseous  abundances of elements are defined, the molecular abundances are determined from the adopted CO:CO$_2$:CH$_3$OH:CH$_4$, and N$_2$:NH$_3$ gas phase molecular ratios. The remaining O gives the abundance of H$_2$O. We set CO:CO$_2$:CH$_3$OH:CH$_4$ = 10:4:1.67:1 in the gas phase of the disk, values consistent with interstellar medium (ISM) measurements \citep{2006A&A...453L..47P,2011ApJ...743L..16O} and measurements of production rates of molecules in Comet C/1995 O1 Hale-Bopp \citep{2004come.book..391B}. In addition, S is assumed to exist in the form of H$_2$S, with H$_2$S:H$_2$ = 0.5 $\times$ (S:H$_2$)$_\odot$, and other S-rich refractory components \citep{2005Icar..175....1P}. We finally consider N$_2$:NH$_3$ = 10:1 in the nebula gas-phase, a value predicted by PSN chemical models \citep{1980ApJ...238..357L}.  

\begin{figure}[h]
\begin{center}
\resizebox{\hsize}{!}{\includegraphics[angle=-90]{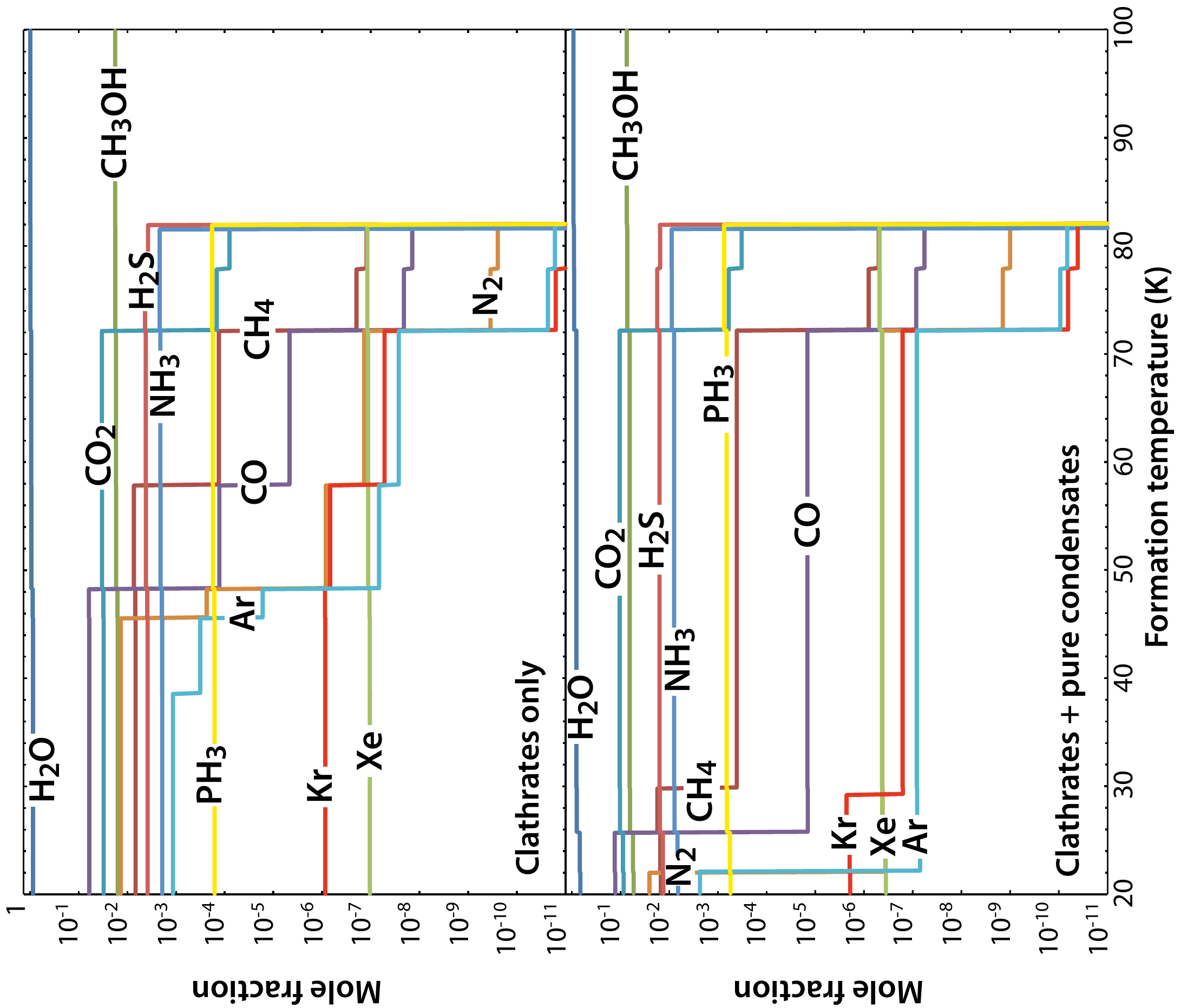}}
\caption{Composition of the volatile phase incorporated in planetesimals formed beyond the snow line in the PSN as a function of their formation temperature. Top:  full clathration scenario. Bottom: limited clathration scenario.}
\label{comp}
\end{center}
\end{figure}

Figure \ref{comp} shows two cases for the compositions of planetesimals condensed in Saturn's feeding zone and represented as a function of their formation temperature. In both cases, NH$_3$ forms NH$_3$-H$_2$O hydrate and CH$_3$OH is assumed to condense as pure ice in the PSN because of the lack of thermodynamic data concerning its associated clathrate. In the first case (full clathration), all volatiles (except NH$_3$ and CH$_3$OH) are trapped in the clathrate phase as a result of an initial supersolar oxygen abundance ($\sim$2.6 $\times$ (O/H)$_{\odot}$) corresponding to H$_2$O/H$_2$ = 2.47 $\times$ 10$^{-3}$ in Saturn's feeding zone. In the second case (limited clathration), we used a protosolar abundance for oxygen, corresponding to  H$_2$O/H$_2$ = 5.55 $\times$ 10$^{-4}$ in Saturn's feeding zone, and implying that the budget of available crystalline water is not sufficient to trap all volatiles in clathrates. In this case, significant fractions of CO, N$_2$ and Ar form pure ices if the disk cools down to very low temperatures ($\sim$20 K), instead of being trapped in clathrates, as it is the case for full volatile clathration. For example, Ar, N$_2$ and CO become substantially trapped in the clathrate phase at $\sim$38, 45, and 48 K in the PSN, respectively. In contrast, these species form pure ices in the 22--26 K range in the PSN.

Assuming that the composition of the icy phase of planetesimals computed with our model is representative of that of Saturn's building blocks, the precise adjustment of their mass accreted by the forming Saturn and vaporized into its envelope allows us to reproduce the observed volatile enrichments. Here, because of the lack of reliable measurements, our fitting strategy is to match the minimum carbon enrichment measured in Saturn. By doing so, this allows us to maintain the mass of solids accreted into Saturn's envelope as small as possible in order to be compared to the mass of heavy elements predicted by interior models.

\section{Results}
\label{resultats}

\begin{figure}[h]
\begin{center}
\resizebox{\hsize}{!}{\includegraphics[angle=0]{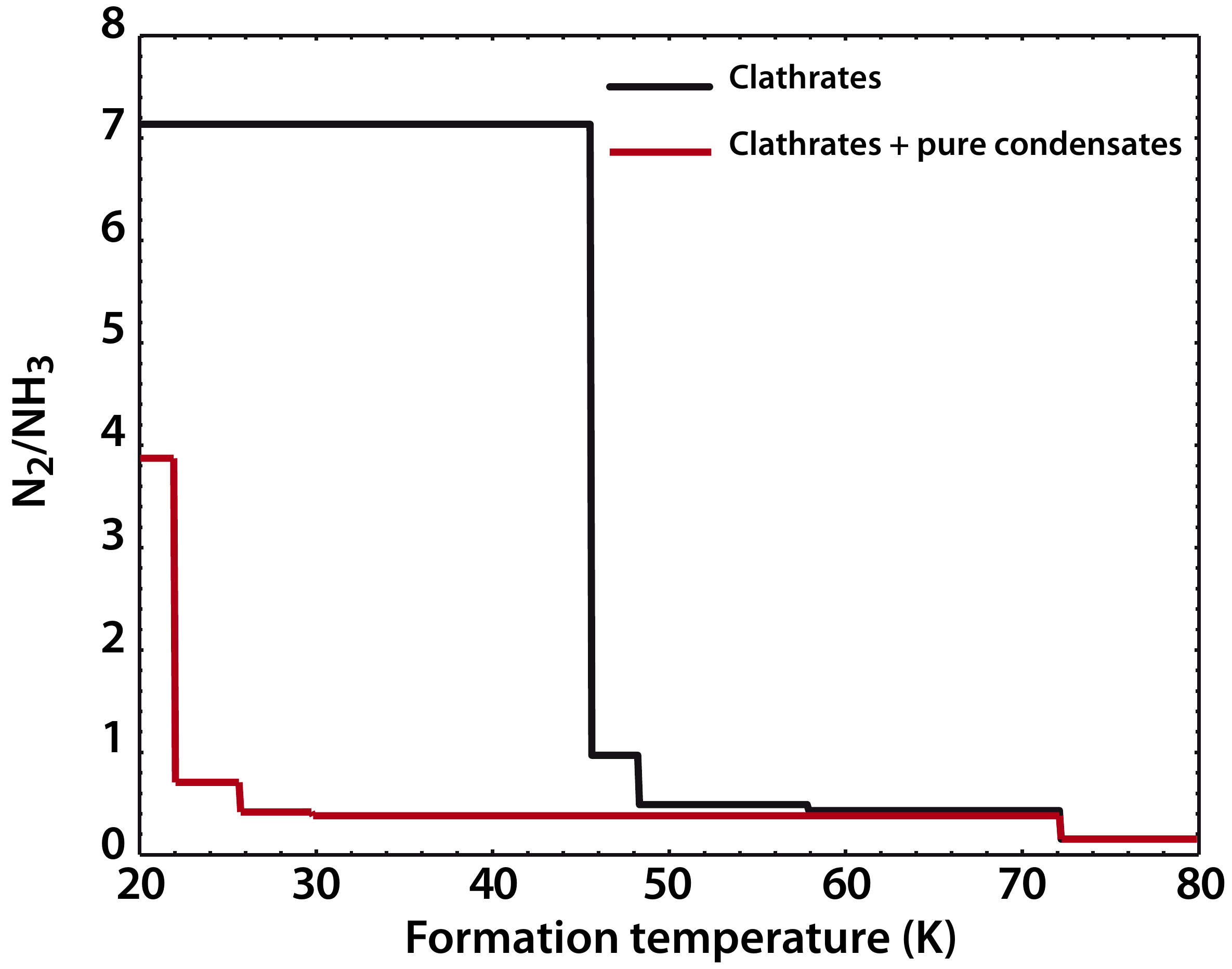}}
\caption{N$_2$/NH$_3$ ratio in the envelope of Saturn as a function of the formation temperature of its building blocks in the cases of full clathration (black curve) and limited clathration (red curve) scenarios.} 
\label{ratio}
\end{center}
\end{figure}

Figure \ref{ratio} shows the evolution of the N$_2$/NH$_3$ ratio in Saturn as a function of the formation temperature of its building blocks, and assuming that it was equal to 10 in the PSN prior to planetesimals formation. Depending on the temperature considered for Saturn's formation, contributions of both N$_2$ and NH$_3$ in solid and gaseous phases have been taken into account in our computation. When not trapped or condensed, the species collapse with the nebula gas onto the forming planet and form its gaseous envelope. NH$_3$ is always in solid form at $T$ $<$ 80 K. In the full clathration case, the maximum temperature of Saturn's formation yielding N$_2$ $\gg$ NH$_3$ in the envelope is $\sim$45 K. Above this temperature, N$_2$ remains essentially in gaseous form. In contrast, in the limited clathration case, N$_2$ dominates in Saturn only at formation temperatures lower than $\sim$22 K, a value corresponding to its condensation temperature in the PSN. In both situations, the amount of N$_2$ supplied to Saturn in gaseous form is less than that of NH$_3$ in solid form. A comparison between the two cases shows that the full volatile clathration favors a higher N$_2$/NH$_3$ in Saturn at temperatures below $\sim$45 K. Because i) disk's temperatures as low as $\sim$22 K at the formation location of Saturn are unlikely and ii) the full clathration scenario is fully consistent with a high N$_2$/NH$_3$ in Saturn, only this latter case is considered in the following.

\begin{figure}[h]
\begin{center}
\includegraphics[angle=0,width=14cm]{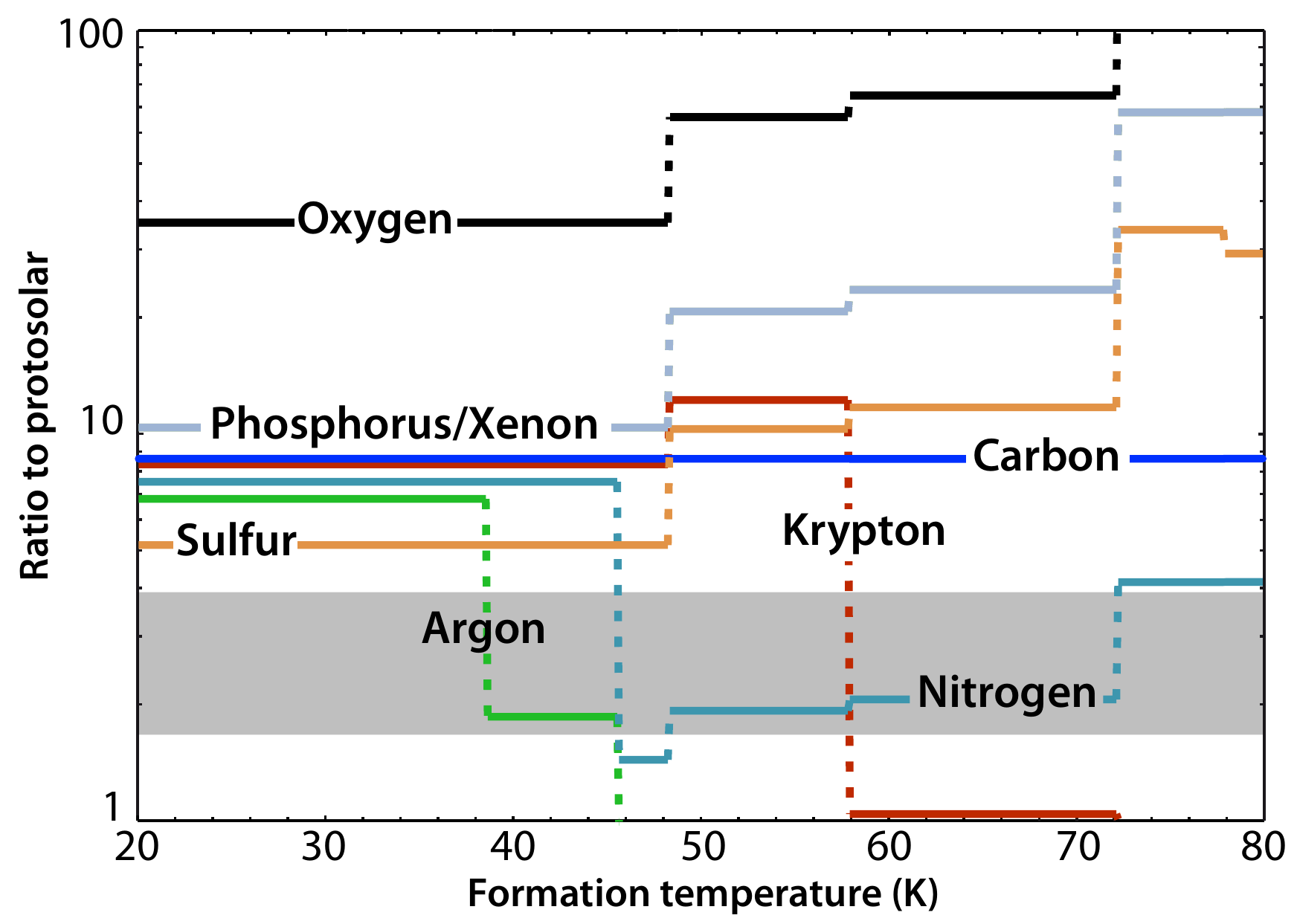}
\caption{Volatile enrichments computed in Saturn's atmosphere as a function of the formation temperature of its building blocks (full clathration scenario). The results have been fitted to the minimum value of carbon enrichment measured in Saturn's atmosphere (see Table 1). P and Xe enrichments appear superimposed and the grey area represents the uncertainties on the N measurement.} 
\label{enri}
\end{center}
\end{figure}

Figure \ref{enri} represents the volatile enrichments in Saturn calculated from the fit of the minimum C abundance observed in the atmosphere ($\sim$8.6 times protosolar -- see Table 1) as a function of the formation temperature of its building blocks and in the case of the full clathration scenario. In the PSN temperature range ($T$ $\le$45 K) consistent with the $^{14}$N/$^{15}$N constraint, we find that N is 7.5 times more enriched than the protosolar value in Saturn's atmosphere, a value higher than the maximum inferred one (3.9 times protosolar), but still lower than the measured C enrichment (9.6 $\pm$ 1 times protosolar). In this case, O is predicted to be at least $\sim$34.9 times more enriched than the protosolar value in Saturn's envelope. This strong O enrichment is due to the assumption of a supersolar abundance of oxygen in Saturn's feeding zone, which is required for the full trapping of guest molecules in clathrates ($\sim$6 water molecules are needed to stabilize one guest molecule). Interestingly, the calculated P enrichment ($\sim$10.4 times protosolar) matches the measured value (11.2 $\pm$ 1.3 times protosolar). S is found 5.2 times more enriched than the protosolar value, but remains lower than the indirect determination ($\sim$12 times protosolar). Table 1 summarizes the enrichments calculated from the minimum and maximum fits of the C abundance observed in Saturn and gives predictions for the volatile species that have not been yet detected (Ar, Kr, Xe) or those whose sampling still needs to be investigated (O, N, S, P). The observed P abundance is matched by our model when C is ranged between 8.6 and 10.4 times protosolar.

\begin{figure}[h]
\begin{center}
\resizebox{\hsize}{!}{\includegraphics[angle=0]{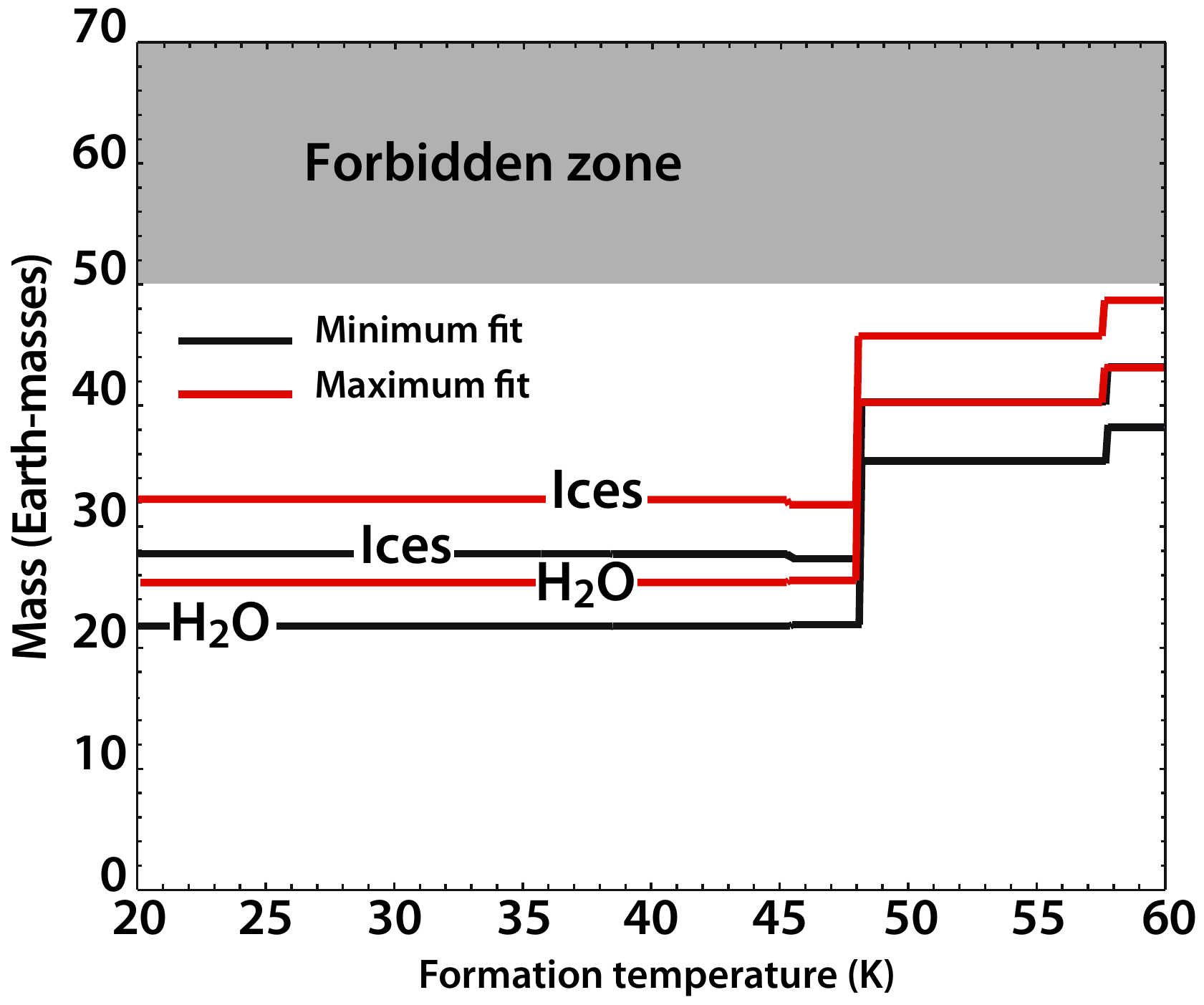}}
\caption{Masses of ices and water needed to be accreted in Saturn's envelope in order to match the minimum and maximum fits of carbon measurement as a function of the planet's formation temperature (full clathration scenario). The forbidden zone corresponds to masses of heavy elements above the maximum value predicted by Leconte \& Chabrier (2012).}
\label{mass}
\end{center}
\end{figure}

Figure \ref{mass} shows that $\sim$27.9--32.2 \Mearth~of ices, including $\sim$21.9--25.3 \Mearth~of water, are needed in Saturn to match the measured C enrichment in the full clathration scenario at $\sim$45 K in the PSN. Our calculated mass range must be seen as a minimum because planetesimals may harbor a significant fraction of refractory phase. We find that 15.2--17.6 \Mearth~of rocks are needed to match the observed C enrichment, assuming an ice--to--rock ratio of $\sim$1 for a protosolar composition gas (Johnson et al. 2012). This implies that  $\sim$43.1--49.8 \Mearth~of heavy elements have been delivered to Saturn to match the observed C enrichment. The mass of heavy elements needed by the full clathration scenario then exceeds the maximum mass of heavy elements predicted in Saturn by homogeneous interior models ($\le$30 \Mearth; \cite{2013Icar..225..548N}). On the other hand, our calculations are consistent with the mass range (26--50 \Mearth) of heavy elements predicted by semi-convective models \citep{2012A&A...540A..20L}.

\section{Discussion}
\label{discu}

The lower limit for the $^{14}$N/$^{15}$N ratio found by \cite{2014Icar..238..170F} implies that Saturn's nitrogen was essentially accreted in N$_2$ form at its formation time. However, this condition is not sufficient to match the measured $^{14}$N/$^{15}$N ratio: our calculations suggest that N$_2$ must have been accreted in solid form in Saturn, in order to match the observed C enrichment, otherwise NH$_3$ would still remain the main N--bearing reservoir in the envelope. 

We have explored two hypotheses to simultaneously account for the $^{14}$N/$^{15}$N measurement and the volatile enrichments in Saturn by varying the O/H ratio in the giant planet's feeding zone. Both possibilities were investigated by using a statistical thermodynamic approach allowing us to compute the composition of clathrates formed in the PSN. In the first case (full clathration scenario), we assumed that oxygen was sufficiently abundant ($\sim$2.6 $\times$ (O/H)$_{\odot}$) to trap all volatiles as clathrates in Saturn's feeding zone (except NH$_3$ which forms a stochiometric hydrate and CH$_3$OH due to the lack of thermodynamic data concerning its associated clathrate), leading to N$_2$ trapping in planetesimals at $\sim$45 K in the PSN. In the second case (limited clathration scenario), we assumed that the O abundance was protosolar, implying that planetesimals were agglomerated from a mixture of clathrates and pure condensates. The PSN had to cool down to $\sim$22 K in order to allow the trapping of solid N$_2$ in planetesimals. The full clathration hypothesis is the only scenario allowing the formation of Saturn's building blocks at temperatures consistent with our knowledge of the thermal structure of the PSN. The presence of a supersolar oxygen abundance in the giant planet's feeding zone may be explained via its formation in the neighborhood of Jupiter, close to the water ice line location at earlier epochs of the PSN. At this location, the abundance of crystalline water ice may have been enhanced by diffusive redistribution and condensation of water vapor \citep{1988Icar...75..146S,2014ApJ...785..125A,2014ApJ...793....9A}, thus easing the formation of clathrates. Recent volatile distribution models, elaborated by \cite{2014ApJ...785..125A,2014ApJ...793....9A} and taking into account the major dynamical and thermodynamic effects relevant to volatiles (turbulent gas drag, sublimation of solids, gas diffusion and condensation), predict enhancements of the surface density of water up to several times the one derived from a protosolar O abundance at the location of the H$_2$O ice line. If the region of the disk where the surface density of water is enhanced extends over several AU, then Jupiter and Saturn could form independently. This would require that the diffused H$_2$O vapor condenses over large length scales, as shown by the simulations of \cite{2013A&A...552A.137R}. Alternatively, recent disk models including midplane deadzone related effects \citep{2012MNRAS.425L...6M,2013MNRAS.432.1616M} show that the temperature profile might not be monotonic in the PSN. These models suggest the occurrence of several ice lines for the same species at some stages of the PSN evolution. In this context, Jupiter and Saturn may have formed at the locations of two distinct water ice lines, each of them providing enough crystalline water for enabling the full clathration of the other volatiles when the giant planets feeding zones reached lower temperatures. In any case, the building blocks accreted by Jupiter and Saturn during their formation should have close C/O and C/N ratios, given the similarity of their formation conditions and the bulk composition of the two planets should also reflect these ratios.

The full clathration hypothesis matches well the formation scenario of Jupiter proposed by \cite{2001ApJ...550L.227G}, where the authors also proposed that the volatiles were fully trapped in clathrates and found O/H $\sim$2.5 $\times$ (O/H)$_{\odot}$ in the giant planet's feeding zone from a simple clathrate formation model. A higher abundance of water ice in Saturn's feeding zone increases the ice-to-rock ratio in planetesimals and implies that the icy phase is dominant in the heavy elements accreted by Saturn's envelope. Given that the full clathration scenario bet fits the data, it favors the idea that Saturn's interior is heterogeneous and may exhibit a continuous compositional gradient, as suggested by \cite{2012A&A...540A..20L}. In order to match this model, one needs to argue that a fraction of the heavy elements sedimented onto Saturn's core during its evolution \citep{2003Icar..164..228F,2004ApJ...608.1039F}. For example, if all rocks sedimented onto Saturn's core ($\sim$15.2--17.6 \Mearth), then the mass of volatiles remaining in the envelope ($\sim$27.9--32.2 \Mearth) holds well within the mass range of heavy elements (10--36 \Mearth) predicted by the semi-convective models of Leconte \& Chabrier (2012). Given the high O enrichment predicted in Saturn (34.9 times the protosolar value), one should expect an increase relative to the D/H ratio in the envelope's hydrogen, a prediction that is not supported by existing observations \citep{2001A&A...370..610L,2003DPS....35.5001B}. However, if the interior of Saturn is semi-convective, the D/H ratio measured in its upper layers would not be representative of the planet's global value.

Interestingly, our results are consistent with the fact that NH$_3$ must be the main primordial reservoir of nitrogen in Titan to explain its current $^{14}$N/$^{15}$N ratio \citep{2014ApJ...788L..24M}. Indeed, formation scenarios predict that Titan's building blocks must have experienced a partial devolatilization during their migration in Saturn's subnebula, which would have induced the loss of the CO, N$_2$ and Ar captured from the nebula \citep{2009ApJ...691.1780M}. Hence, Titan's building blocks probably originate from Saturn's feeding zone but they would have been subsequently altered by the subnebula. Finally, notwithstanding the conclusions of the present study, it should be kept in mind that only the {\it in situ} measurement of O below the condensation layer of water, and the precise assessment of the C, N, P and the noble gas abundances will be able to shed light on the formation conditions of the ringed planet \citep{2014arXiv1404.4811M}. 

\acknowledgements
O.M. acknowledges support from CNES. J.I.L. thanks the Juno project for its support of his work. L.N.F. was supported by a Royal Society Research Fellowship at the University of Oxford. This work has been carried out thanks to the support of the A*MIDEX project (n\textsuperscript{o} ANR-11-IDEX-0001-02) funded by the ``Investissements d'Avenir'' French Government program, managed by the French National Research Agency (ANR).

%% ------------------------------------------------------------------------ %%
%
%  REFERENCE LIST AND TEXT CITATIONS
%
% Either type in your references using

\clearpage

\begin{table}[h]
\begin{center}
\caption[]{Observed and calculated enrichments in volatiles in Saturn}
\small{\begin{tabular}{lcc}
\hline
\hline
\noalign{\smallskip}
Species			& Measurements$^{(\star)}$					& Minimum and maximum fits of C 	\\
\noalign{\smallskip}
\hline
\noalign{\smallskip}
O				& --										& 34.9 -- 43.0					\\
C				& 9.6	 $\pm$ 1.0$^{(a)}$						& 8.6	 -- 10.6					\\
N				& 2.8 $\pm$ 1.1$^{(b)}$						& 7.5 -- 9.2					\\
S				& 12.05$^{(c)}$								& 5.2 -- 6.4					\\	
P				& 11.2 $\pm$ 1.3$^{(d)}$						& 10.4 -- 12.7					\\	
Ar				& --										& 1.9	 -- 2.3					\\
Kr				& --										& 8.3	 -- 10.3					\\
Xe				& --										& 10.4 -- 12.7					\\
\noalign{\smallskip}
\hline
\end{tabular}}\\
Notes. Saturn's formation temperature is considered at $\sim$45 K. The observed values are derived from $^{(a)}$Fletcher et al. (2009a), $^{(b)}$Fletcher et al. (2011), $^{(c)}$Briggs \& Sackett (1989) and $^{(d)}$Fletcher et al. (2009b), using the protosolar abundances of \cite{2009LanB...4B...44L}. $^{(\star)}$Error is defined as ($\Delta$E/E)$^2$~=~($\Delta$X$_{Saturn}$/X$_{Saturn}$)$^2$ + ($\Delta$X$_\odot$/X$_\odot$)$^2$.
\end{center}
\label{data}
\end{table}

\clearpage

\clearpage

\clearpage

\clearpage

\end{document}